# Self-powered sensors enabled by wide-bandgap perovskite indoor photovoltaic cells

Ian Mathews, Sai Nithin Reddy Kantareddy, Shijing Sun, Mariya Layurova, Janak Thapa, Juan-Pablo Correa-Baena, Rahul Bhattacharyya, Tonio Buonassisi, Sanjay Sarma and Ian Marius Peters.
Department of Mechanical Engineering, Massachusetts Institute of Technology, Cambridge, MA 02139, USA

**Abstract**

We present a new approach to ubiquitous sensing for indoor applications, using high-efficiency and low-cost indoor perovksite photovoltaic cells as external power sources for backscatter sensors. We demonstrate wide-bandgap perovskite photovoltaic cells for indoor light energy harvesting with the 1.63eV and 1.84 eV devices demonstrate efficiencies of 21% and 18.5% respectively under indoor compact fluorescent lighting, with a champion open-circuit voltage of 0.95 V in a 1.84 eV cell under a light intensity of 0.16 mW/cm$_2$. Subsequently, we demonstrate a wireless temperature sensor self-powered by a perovskite indoor light-harvesting module. We connect three perovskite photovoltaic cells in series to create a module that produces 14.5 μW output power under 0.16 mW/cm$_2$ of compact fluorescent illumination with an efficiency of 13.2%. We use this module as an external power source for a battery-assisted RFID temperature sensor and demonstrate a read range by of 5.1 meters while maintaining very high frequency measurements every 1.24 seconds. Our combined indoor perovskite photovoltaic modules and backscatter radio-frequency sensors are further discussed as a route to ubiquitous sensing in buildings given their potential to be manufactured in an integrated manner at very low-cost, their lack of a need for battery replacement and the high frequency data collection possible.

**Introduction**

Deploying wireless sensors at scale will generate the Big Data required to optimize operations and increase efficiency across many industries primarily located indoors such as energy-efficient and smart buildings [1], logistics and inventory sensing and tracking [2], health monitoring [3], or robotic systems [4], [5]. In order to widely deploy wireless sensors for indoor applications, we investigate here the manufacture of devices that are low-cost, self-powered, easily deployable at scale and could remain operational for many years. To manufacture self-powered sensors that satisfy these criteria, a holistic approach to hardware design is required. The three core hardware components include the power supply, communication components and sensor(s). We are investigating the low-cost and integrated manufacture of these components, for example, by printing a photovoltaic cell and radio-frequency (RF) antenna on the same flexible substrate. As a first step, in this paper, we investigate perovskite photovoltaic cells fabricated on glass as potential low-cost power sources for our device, and demonstrate them powering a self-powered RF-backscatter temperature sensor by harvesting ambient light.

Perovskite solar cells have recently revolutionized the field of thin-film photovoltaics as an emerging low-cost solar technology with a record power conversion efficiency (23.7%) now exceeding polycrystalline Si cells (22.3%) [6]. Prototypical lead halide perovskites have a chemical fomula of $APbX_3$ ($A$ = methylammonium (MA), formamidinium (FA); and $X$ = Br, I) while more recently, mixed cation perovskites which incorporate Cs and Rb have demonstrated improved photovoltaic performance and device stability [7], [8]. Perovskites have the potential to be manufactured at very low-cost through solution processing [9], [10], that could also enable simple and integrated IoT device fabrication where an indoor photovoltaic (IPV) cell is printed on the same plastic substrate as the RF antenna and sensor, as outlined in Fig 1 (b).

Indoor light-harvesting perovskite cells with conversion efficiencies of >25% under ambient lighting have recently been demonstrated [11], [12]. Both examples use a perovskite composition with optical bandgaps of ~1.6 eV, lower than the 2 eV optimum for conversion of low-intensity illumination from compact-fluorescent (CFL) or white-LED sources — a value at which IPV cell efficiencies of over 50% are theoretically possible [9]. The enormous chemical diversity of hybrid organic-inorganic perovskites allows tuning of optoelectronic properties where bandgaps between 1.5 eV to 2.2 eV have been demonstrated by altering the Br to I ratio in lead halide perovskites [13], [14]. Therefore, in this work we fabricate and test wide-bandgap perovskite cells (>1.6 eV) for indoor light-harvesting to provide a potential boost in cell efficiency. In addition to an increase in efficiency, using wide-bandgap perovskite IPV cells should result in higher operating voltages, reducing the need for inefficient DC-DC power conversion electronics to boost the cell voltage to that required by the wireless device [15].

For a truly integrated approach to designing and manufacturing self-powered sensors, suitable low-power communication and sensing components must also be considered. For indoor wireless sensing, (RF) backscatter has recently emerged as an extremely low-cost technology that achieves significant range while consuming only μW's of power for computation. Fig. 1 (a) compares the cost and power consumption of a number of demonstrated communication systems for indoor wireless sensors using data from [16]. It is clear that protocols that rely on backscatter consume much less power than active radio technologies, a characteristic achieved by modulating and reflecting the RF interrogator signal rather than generating an RF response signal using an onboard radio. This is, however, at the expense of communication range and data-rate where active radio methods can send more data over longer distances. Nevertheless, backscatter-based communication techniques have recently been shown to work indoors with measured communication ranges of up to 10's of meters in buildings, including through multiple walls for ultra-high-frequency [16], [17]. Moreover, the sensitivities of a passive RF tag's integrated chip (IC) improve by 1 dBm per year enhancing communication range over time.

In typical passive backscatter systems, incident RF energy is harvested to power-on the IC and an internal RF switch modulates the IC's load impedance to encode bit data on the backscattered RF signal. Tag ICs are typically power hungry, so the communication range is forward link limited, i.e., if the chip receives enough power to turn on, it is virtually guaranteed to be read. The range of the backscatter signal, however, is greatly increased by using a coin cell battery or small IPV cell to provide the additional power to handle computations, reducing the dependence on the RF signal to power the IC. Therefore, the maximum RF signal is backscattered to the reader, increasing read range [18]. This presents an opportunity for well-designed IPV cells to become an integral part of RF backscatter sensing systems.

In the rest of this paper, we begin by reporting the photovoltaic performance of wide-bandgap IPV cells made from $(Rb_{0.01}Cs_{0.05})(MA_xFA_{1-x})_{0.94}Pb(Br_xI_{1-x})_3$ perovskite materials under ambient light conditions [19]. We vary the I-Br composition to produce a 1.84 eV device, close to the optimum for IPV cells. We report the high efficiency (18.5%) and high photo-voltage (0.95 V) measured under ambient illumination. We subsequently demonstrate a low-cost self-powered environmental sensor by using three wide-bandgap perovskite IPV cells to power an RFID Ultra-High-Frequency (UHF) integrated chip with an onboard temperature sensor, connected to a laser-cut copper antenna. Compared to a standard backscatter RFID tag sensor without an external power source, we demonstrate a maximum 7.2x boost in communication range, achieved by reducing the amount of RF power harvested that is used to power the tag's IC, using the external power supplied by the perovskite IPV cells instead, such that the maximum RF signal is backscattered. The combination of these components represents our first step towards truly integrated manufacture of self-powered sensors.

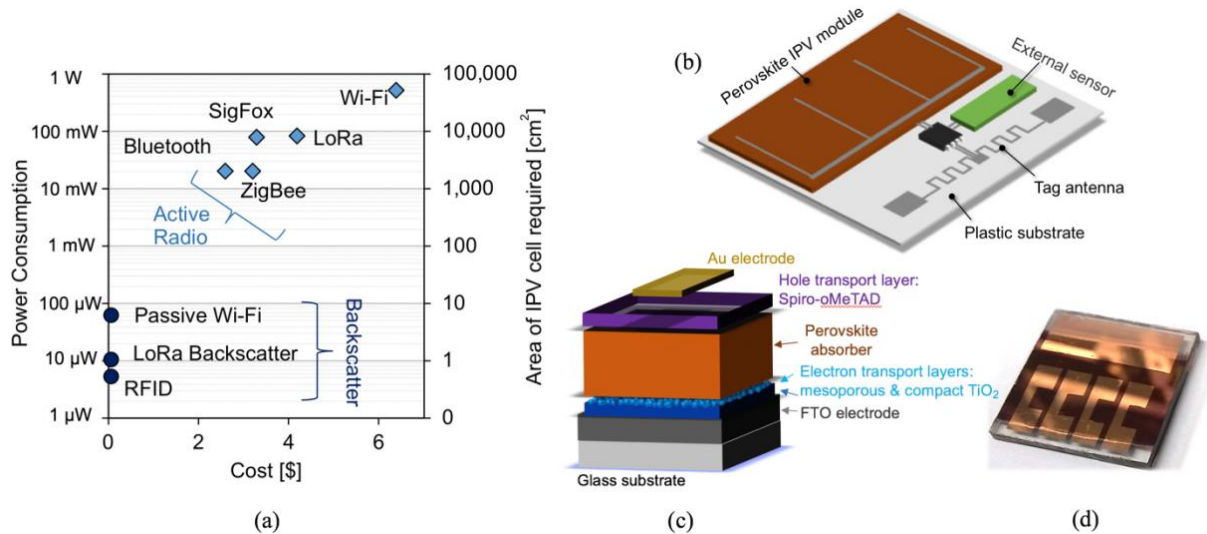

Figure 1: (a) The power consumption and cost of multiple active radio and backscatter RF communication protocols as given in [16] (b) a schematic layout of our proposed IPV-backscatter sensors (expected substrate area of 5-10 cm$^2$) (c) an exploded view of the layers in our perovskite photovoltaic cell and (d) an image of a fabricated perovskite device with 4 individual cells.

**Results & Discussion**

*(i)    Cell characterization*

We compare the performance of two perovskite absorbers: (i) our standard recipe which adapts the state-of-the-art high-performance perovskite solar cells with a composition of $(MA_{0.17}FA_{0.83})Pb(Br_{0.17}I_{0.83})_3$ and (ii) a wider bandgap solution of $(MA_{0.5}FA_{0.5})Pb(Br_{0.5}I_{0.5})_3$. In both compositions, 1% RbI and 5% CsI were added in the precursor solution to enhance the device stability [19]. Before fabricating cells, we measured the absorbance of the perovskite films and determined optical band gaps of 1.63 eV and 1.84 eV for the standard and wide-bandgap films, respectively. Solar cells with these perovskite absorber layers were fabricated with a structure as shown in Figures 1 (c) & (d) with full fabrication details provided in the methods section. EQE measurements for both cell types, taken under an illumination bias of 1000 W/m$_2$, confirmed that they exhibit strong carrier collection in the 400-650 nm range of interest for harvesting light from typical compact fluorescent and white-LED sources, as shown in Fig. 2 (a). The wide-bandgap cell has an EQE of 50-60% in this range, lower than for the standard cell (~70%) indicating that our recipe and/or cell structure are not yet fully optimized. The photovoltaic conversion efficiency of the cells under solar illumination were 16.5% for the standard recipe cell and 11.9% for the wide-bandgap cell with the current-voltage characteristics of our champion cells presented in Fig. 2 (b). A 1 sun efficiency of ~12% is close to the state-of-the-art for 1.8 eV perovskite cells [20]. Using a wider bandgap absorber leads to a 78 mV boost in open-circuit voltage in our cell design with an increase from 1082 mV to 1160 mV under 1 sun illumination.

To investigate these cells' performance as indoor light harvesters, we measured the current-voltage characteristics of the devices under 0.16 mW/cm$_2$ illumination from a compact-fluorescent source as shown in Fig. 2 (c). This equates to standard light levels in an office space without any solar insolation through a window. There is a small difference in the current produced by both cells that we attribute to the lower EQE achieved by the wide-bandgap device across the entire range of interest. The boost in open-circuit voltage from using the wide-bandgap absorber is 72 mV with the best wide-bandgap device having an open-circuit voltage, $V_{OC}$, of 955 mV at this low light intensity. The only cells in the literature that achieve higher values are 1.8 eV GaInP devices that showed open-circuit voltages of 1.2 V under similar low light intensities. Overall, the standard and wide-bandgap cells converted the low intensity indoor light with efficiencies of 21.4% and 18.5% respectively. The theoretical maximum efficiencies of these two devices under low intensity

CFL or LED lighting are ~45% and ~51%, respectively, indicating a greater upside potential for the wider-bandgap device with further optimization.

To further investigate the devices performance under low light intensities we measured the photovoltaic performance under a solar simulator across 5 orders of magnitude light intensity using neutral density filters – (using an AM1.5G spectrum for all measurements). Fig. 2 (d) plots the change in $V_{OC}$ of a 1.84 eV cell versus light intensity, and compares it to the open-circuit voltages achieved for other leading IPV cells made from GaAs and dye-sensitizing materials [21]. The cell used for this measurement represented the average device produced in a batch and while it does not achieve as high a $V_{OC}$ as our champion device, the results show comparable voltages to the best GaAs cells under low light intensities. The ideality of the wide-bandgap cell derived from the slope of the $V_{OC}$ curve is ~1.8, indicating recombination is mostly due to distributed Shockley-Read-Hall recombination via bulk defects [22].

The high open-circuit voltages demonstrated by our wide-bandgap IPV cells show their suitability for IoT nodes where it would require a small number of cells to provide charging voltages of 1.5, 3.3, or 5 V, reducing the need for inefficient power management electronics. Despite our standard recipe perovskite absorber having a marginally higher efficiency under low light conditions, given the extra voltage produced by the 1.86 eV cells under ambient light, we proceed in the following section to use a number of wide-bandgap devices connected in series to form an IoT module to power self-powered temperature sensors.

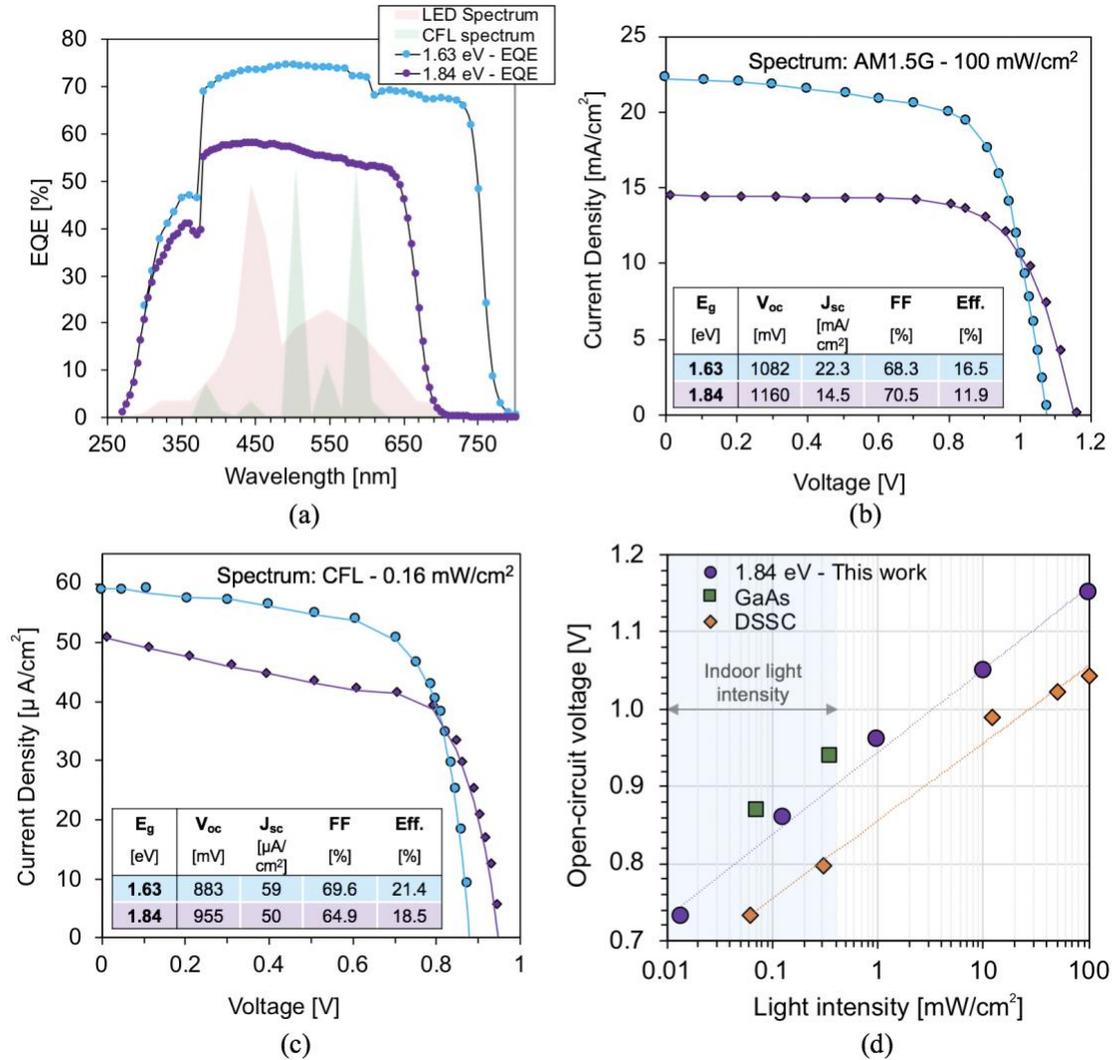

Figure 2: (a) The measured EQE of both cell types compared to standard CFL and white-LED spectra, (b) the current-voltage behavior of both cell types under 1 sun illumination, (c) the current-voltage behavior of both cell types under 0.16 μW/cm2 CFL illumination and (d) the measured open-circuit voltage of the 1.84 eV cell as a function of light intensity and compared to values for dye-sensitized (DSSC) and GaAs cells optimized for indoor conditions [21] that had ambient light-harvesting efficiencies of 29% and 21% respectively[1].

*(ii)    Self-powered temperature sensor*

---

[1] (b) & (c) See Supplementary Information for forward and reverse voltage sweeps

In this section we present our first IPV-backscatter device and compare its communications range to a standard RFID sensor. Our initial prototype consists of our 1.84 eV wide-bandgap IPV cells integrated into an IPV module with 3 cells connected in series, an RFID IC with an internal temperature sensor and a laser-cut Cu antenna with the circuit diagram for this prototype provided in Fig. 3 (a) and images of the fabricated device in Figs. 3 (e)&(f). The device operates by drawing power from the cell to power on-board electronics. Data however is conveyed by backscatter as is typical in battery-assisted passive devices. No energy storage is provided in our initial prototype, to do so would require a larger IPV module to charge an on-board capacitor of supercapacitor to provide power when operating in the dark as in our previous work with a-Si PV cells [23]. By connecting our prototypes in this manner, we avoid the need for power management circuitry reducing the cost and complexity of our design.

Theoretically, a communication range of 40-50 m range is possible with semi-passive RFID tags powered by an external power source [23]. The upper limit is set by the sensitivity of the readers and RFID IC, and the maximum transmitted power allowed by the FCC. In practice the read-range will be limited by many factors including the antenna gain, antenna-IC impedance matching, transmission power, multi-path interference, contact losses, background dielectric, etc. The optimum design of an RFID antenna for indoor applications is beyond the scope of this paper. Instead, we are interested in characterizing the increase in range possible when powering the device using external IPV cells. This scalar increase will be similar to the increase possible for any semi-passive antenna design, including optimized ones.

Our wireless sensor design relies on the IPV module to provide the required power to turn-on the IC, providing much greater range than in the case where part of the incoming RF signal is harvested for this purpose. For this mode of operation to work, an IC is required that can switch between RF harvesting or external power mode. We use the EM4325 from EM Microelectronics, a Class-3 Generation-2 UHF RFID IC. To operate this IC using an external power source requires an input voltage of >1.5 V and an input power of >10 µW. To provide these power levels from perovskite PV cells under ambient light we mechanically separated individual cells from the glass substrate and connected 3 in series using external wiring. The 3 cells combined had an active area of 0.7 cm$_2$ as defined by the Au contact pads. No masking was used during testing of the module with the measured current-voltage and power-voltage curves of the indoor perovskite module under 0.16 mW/cm$_2$ compact-fluorescent illumination shown in Fig. 3(b). The module had an open-circuit voltage of 2.62 V (870 mV per cell) indicating some losses in voltage occur during module packaging. The short-circuit current produced by the 3 cells in series was 10.8 µA or ~46 µA/cm$_2$. The maximum power output from the module of 14.5 µW at 1.81 V is more than enough to provide external power to the RFID IC and operate the device autonomously when illuminated by ambient light. This small module had an efficiency of 13.2% under these low lighting conditions.

The communication range of individual tags is measured using a Voyantic Tagformance Pro with the tag held 30 cm from the interrogator antenna [24]. To confirm the range increase achieved by using the perovskite IPV module as an external power source, we measured the power backscattered by the tag when powered by the module, illuminated by a CFL light source, and a tag without any external power source. We measured over an 800–1000 MHz range covering the global allowable frequency ranges for RF devices. The communication range of our RFID tags without an external power source is 0.7-0.95 meters within the 800–1000 MHz frequency range (Fig. 3 (c)). Within the same range, a range increase of 4x–7.2x was measured when using the IPV module as an external power source, with a max absolute range of 5.1 meters. While this tag design does not have an absolute range large enough to allow one reader to communicate with multiple RFID sensors spread out over a building, *e.g.*, in a person's home, optimization of the antenna, increased power from the reader or switching to LoRa backstatter protocols can all be used to increase the absolute range to practical levels without a significant increase in cost.

As a confirmation of the efficacy of our wireless temperature sensor, we measured the temperature in our laboratory over a period of ~8 hours. As shown in Fig. 3 (d), our tag sensor setup resulted in 23625 measurements over a period of 29454 seconds or a measurement every 1.24 seconds. These high frequency measurements are possible because our design is not dependent on waiting for a supercapacitor or battery to charge up after every measurement; instead current is drawn from the IPV module to boost the backscatter signal as required. Furthermore, the RFID reader manufacturer's claim maximum throughputs exceeding 1000 tags per second, whereas practical throughput is around 60 tags per second (depending on the total population of tags in the environment and the reader parameters). We find the IPV-backscatter device is able to communicate every few seconds, allowing the collection of 1000's of data points in an hour. This is a two order of magnitude increase in data collection rate compared to other indoor-light harvesting sensors [25] and can undoubtedly provide the high-resolution data required to train machine learning and artificial intelligence algorithms for many applications. While the time constant for temperature change in a room might not require such a high sampling rate, we measure temperature here to show that high sampling rates are possible. Examples where 1000s of data points might be needed in indoor settings include occupancy sensors, logistics and inventory tracking, or positional sensors for manufacturing lines or other robotic systems.

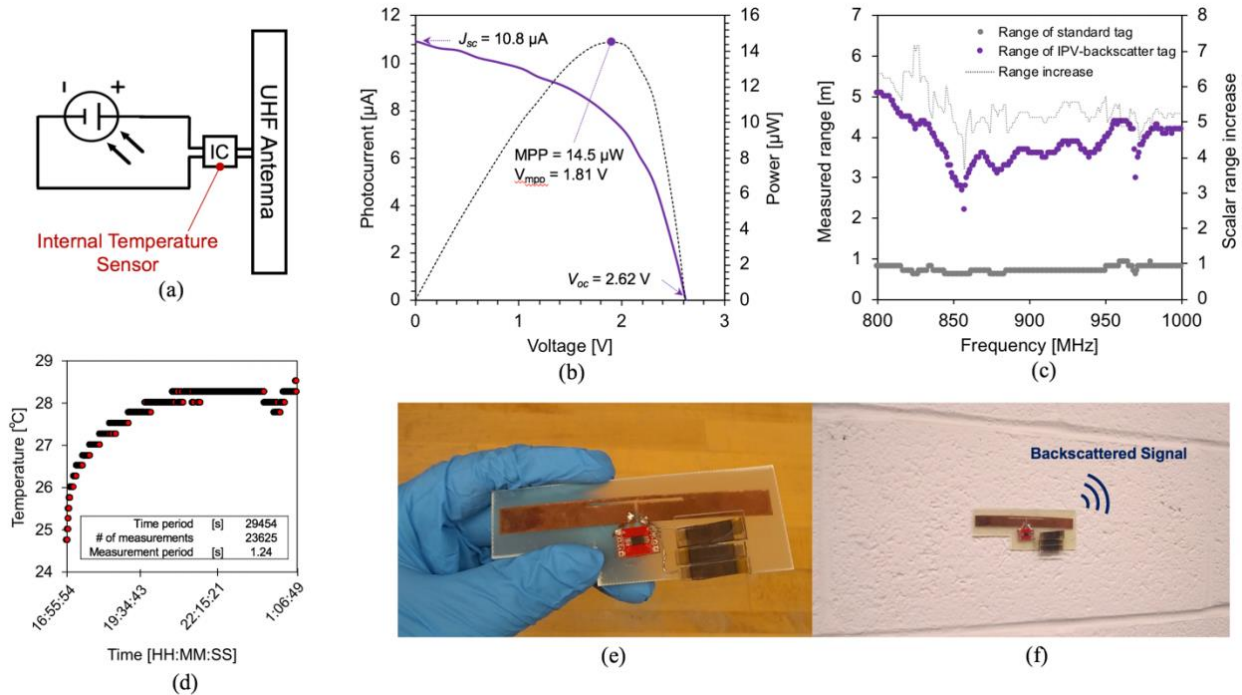

Figure 3: (a) A circuit diagram for the prototype IPV-backscatter sensor, (b) the measured current-voltage and power-voltage output of the indoor perovskite module under 0.16 $\mu$W/cm$_2$ compact fluorescent illumination used to power the sensor tag, (c) the measured range and range increase for a standard and IPV powered sensor tag (d) temperature measurements taken by the sensor tag over a period of ~ 8 hours and (e)&(f) pictures of the of the IPV-backscatter sensor.

**Conclusions**

We presented a new approach to enable ubiquitous sensing in buildings. Considering the requirement for multiple-meter communications range, operation over a number of years, and high-frequency measurements, all at a suitable price-point, we have identified the combination of suitably stable indoor perovskite photovoltaic cells, RF backscatter communication modules and low-cost sensors as an approach that could satisfy these criteria.

We demonstrated wide-bandgap (1.84 eV) perovskite photovoltaic cells with high efficiencies and photovoltages under ambient light. Our champion cell maintained an efficiency of 18.5% and an open-circuit voltage of 0.95 V under compact fluorescent lighting intensities as low as 0.16 mW/cm$_2$. We used three of our cells connected in series to power an RFID temperature sensor. The 0.7 cm$_2$ IPV module produced 14.5 µW under 0.16 mW/cm$_2$ of compact fluorescent illumination, *i.e.*, the IPV module had an efficiency of 13.2% under these low lighting conditions. This external power source increased the backscatter communication range of the sensor by up to 7.2x, reaching a maximum of 5.1 meters while maintaining a measurement period of 1.24 seconds.

One of the primary advantages of the combination of the technologies we have outlined are their suitability to be fabricated in an integrated manner using printing technologies, potentially allowing extremely low-cost manufacturing of wireless sensors. Printed perovskite photovoltaic cells are expected to be manufacturable at a few cents per cm$_2$ while RF antennas are already industrially printed at such low costs. The only component that cannot be printed is the IC that would need pick-and-place integration. In total, a combination of these elements can produce self-powered indoor sensors at very low cost if manufactured at scale – a requirement for the projected scale for the internet of things to become reality.

**Acknowledgements**

The authors acknowledge the sources of funding for this work. I.M. has received funding from the European Union's Horizon 2020 research and innovation programme under the Marie Skłodowska-Curie grant agreement No. 746516. S.N.R.K. has received funding from GS1 organization through the GS1-MIT AutoID labs collaboration. J.P.C.B. by an EERE DOE postdoctoral fellowship. S.S., J.T. and M.L. were supported by a TOTAL SA research grant funded through MITei and Skoltech as part of the Skoltech NGP Program. I.M.P. was financially supported by the DOE-NSF ERF for Quantum Energy and Sustainable Solar Technologies (QESST) and by funding from Singapore's National Research Foundation through the Singapore MIT Alliance for Research and Technology's "Low energy electronic systems (LEES)" IRG.

**Methods**

*(i) Solar cell fabrication*

For our standard recipe, the perovskite films were deposited following the previous report [7]. Solid FAI (Dyesol), MABr (Dyesol), PbI$_2$ (TCI Chemicals) and PbBr$_2$ (TCI Chemicals) were mixed in anhydrous DMF:DMSO 9:1 (v:v, Acros) with molar ratios of FAI:MABr: PbI$_2$: PbBr$_2$ = 1: 0.2: 1.1: 0.22, we then added 5% CsI and 1% RbI (molar ratio to Pb). For our wide-bandgap recipe, the precursor solution contains molar ratios of FAI:MABr: PbI$_2$: PbBr$_2$ =0.5 ∶ 0.5 ∶ 0.55: 0.55, followed by dopants of 5% CsI and 1% RbI. XRD measurements of both the standard and wide-bandgap films are provided in Figures 6 & 7 in the Supplementary Information.

To prepare the electron-selective layer, 20 mM titanium diisopropoxide bis(acetylacetonate) solution (Aldrich) in ethanol was deposited onto a cleaned patterned-F-doped SnO$_2$ (FTO, Pilkington, TEC8) substrate at 450°C by spray pyrolysis, forming a compact layer of TiO$_2$. Prior to spray pyrolysis, as purchased substrates were cleaned by sonicating in 2% Hellmanex detergent in DI water, followed by ethyl alcohol and acetone for 15 minutes each. Using a diluted TiO$_2$ paste (3:5:1 W/W of terpineol: 2-methoxy ethanol), a mesoporous TiO$_2$ (meso-TiO$_2$, Dyesol) film was then spin-coated onto the TiO$_2$/FTO substrate. To remove the organic solvent, the substrate was heated at 500°C for 1 h. Perovskite precursor solution was deposited with a two-step spin-coating program: 10 s at 1000 rpm, followed by 20 s at 6000 rpm. 150 μL of chlorobenzene was deposited 5 s after the start of the second step. The perovskite films were annealed for 15 mins at 100 °C. Spiro-OMeTAD was employed as the hole transport layer, where the doped spiro-OMeTAD (Merck) solution was spin-coated on the perovskite films for 20 s at 4000 rpm. The spiro-OMeTAD solution contains 227 μL of Li-TFSI (Sigma-Aldrich, 1.8 M in acetonitrile) solution, 394 µL of 4-tert-butylpyridine (Sigma-Aldrich) solution, 98 µL cobalt complex (FK209, Lumtec, 0.25 M tris(2-(1H-pyrazol-1-yl)-4-tertbutylpyridine) cobalt(III) tris(bis(trifluoromethylsulfonyl)imide) in acetonitrile) solution, and 10,938 µL of chlorobenzene for every gram of spiro-OMeTAD. For top electrode, an Au electrode (~100 nm) was deposited by thermal evaporation.

*(ii) Bandgap determination*

Transmission and reflection were measured for as-synthesized thin-film samples using Perkin-Elmer Lambda 950 UV/Vis Spectrophotometer. Absorptance was calculated using $A = 1-T-R$. Using the method established by Tauc,[17] we extracted the band gap for the thin films. Band gaps were calculated for both direct and indirect bandgap assumption. Approximate thicknesses of 300 nm film for 1.2M precursor concentration and 100nm film for 0.6M precursor concentration were used for optical properties estimation. The Tauc plots are provided in the Supplementary Information.

*(iii) Photovoltaic characterization*

The Solar Simulator measurement set up includes an Oriel 3A Class AAA Solar Simulator, including AM1.5G optical filter, designed to simulate the AM1.5G solar spectrum. Current-voltage sweeps are performed by a Keithley 2400. A mono-Si reference cell, calibrated by National Renewable Energy Laboratory's Solar cell/Module Performance Group on the 6th March 2018, is used establish 1 sun light intensity. A temperature control stage maintains samples at 25 °C.

The indoor photovoltaic performance measurement set up use the same electronics and sample stage as for 1 sun measurements, but the cell is illuminated using a 17W 3500K fluorescent bulb. The intensity of this low-level illumination is measured using a calibrated Si photodiode.

The EQE station includes a Xenon lamp (with accompanying power supply). A monochromator and filter wheel assembly isolates chopped light of a specific wavelength while optical mirrors guide the light onto the sample stage. The system measures the quantum efficiency by comparing the current from the device to a calibrated Si photodiode.